\documentstyle[aas2pp4]{article}                

\newcommand{\sy}{{\rm {s\,\,yr^{-1}}}}
\newcommand{\etal}{{\it et al. \/}}

\lefthead{Colpi \etal}
\righthead{THE ELUSIVENESS OF OLD NEUTRON STARS}

\begin{document}

\title{The Elusiveness of Old Neutron Stars} 
\author{Monica Colpi}
\affil{Dept. of Physics, University of Milano, \\ Via Celoria 16, 20133 Milano,
Italy \\ e--mail: colpi@uni.mi.astro.it}
\author{Roberto Turolla}
\affil{Dept. of Physics, University of Padova, \\ Via Marzolo 8, 35131 Padova,
Italy \\ e--mail: turolla@astaxp.pd.infn.it}
\author{Silvia Zane\altaffilmark{1}}
\affil{International School for Advanced Studies, \\ Via Beirut 2-4, 34014
Trieste, Italy \\ e--mail: zane@sissa.it}
\altaffiltext{1}{present address: Nuclear and Astrophysics Laboratory, 
University of Oxford, Keble Road, Oxford OX1 3RH, England}
\and
\author{Aldo Treves}
\affil{II Faculty of Sciences, University of Milano, \\ Via Lucini 3, 
22100, Como, Italy \\ e--mail: treves@uni.mi.astro.it}

\begin{abstract}

Old neutron stars (ONSs) which have radiated
away their internal and rotational energy may still shine 
if accreting the interstellar medium. Despite their large number, only 
two promising candidates have been detected  so far and rather stringent 
limits on their observability follow from the analysis of ROSAT
surveys. This contrasts with  optimistic theoretical 
estimates that predicted a large number of sources in ROSAT
fields. We have reconsidered the issue of  ONSs observability, accounting
for the spin and magnetic field evolution over the neutron star
lifetime. In the framework of a spin--induced field decay model, we show 
that the total number of ONSs which are, at present, in the accretion 
stage is reduced by a factor $\sim 5$ over previous figures if the 
characteristic
timescale for crustal current dissipation is $\sim  10^8-10^9$ yr. This 
brings theoretical predictions much closer to observational limits. 
Most ONSs should be at present in the propeller phase and, if subject to
episodic flaring, they could be observable. 

\end{abstract}


\section{Introduction}
Old neutron stars (ONSs), i.e. neutron stars which have evolved beyond
the pulsar phase, are expected to be quite a large Galactic population
counting as many as  $N\sim 10^9$ objects.
Ostriker, Rees, \& Silk (1970)\markcite{ors70} were the first to suggest 
that accretion of the interstellar medium (ISM) may produce enough luminosity 
to make the closest stars observable and that ONSs may contribute to the 
soft X--ray background.
More recently, Treves, \& Colpi (1991)\markcite{tc91} and Blaes, \& Madau
(1993)\markcite{bm93} have shown that several thousands ONSs should be present
in the ROSAT All Sky Survey. Rather optimistic predictions
for the number of detectable sources were also presented by
Madau, \& Blaes (1994)\markcite{mbl94} and Colpi, Campana, \&
Treves (1993) \markcite{cct93} for ONSs embedded in Giant Molecular Clouds
(GMC) and by Zane \etal (1996) \markcite{zztt96}
for ONSs in the solar proximity. However, despite the intense observational
efforts, the search for ONSs produced, up to now, just a handful of candidates, 
out of which only two, RX J18653.5-3754 (Walter, Wolk, \& Neuh\"auser 
1996\markcite{wwn96})
\markcite{wwn96} and RX J0720.4-3125 (Haberl \etal 
1996\markcite{hab96}, 1997\markcite{hab97}), 
seem indeed promising. Moreover, recent analyses of ROSAT fields 
in the 
direction of GMCs (Belloni, Zampieri, \& Campana 1996\markcite{bzc96}; 
Motch \etal 1997\markcite{mo97})
placed rather stringent upper limits on the number 
of ONSs in ROSAT images. In all these cases, the predicted number of ONSs 
turns out to be too  close to the observed number of 
non optically identified sources (NOIDs) 
which should comprise many other types of objects, like white dwarfs, active 
coronae and AGNs. In the field 
investigated by Motch {\it et al.\/}\markcite{mo97}, for example, 
$\sim 5$ sources should be detectable above 0.02 count$\rm s^{-1}$ 
(Zane \etal 1995\markcite{zan95}) while the number of NOIDs, at the same 
limiting flux, is 8. Since these are sites where
accretion should be the most effective, observations 
seem to suggest that current theoretical predictions are in excess of 
at least a 
factor $\sim$ 5--10. Moreover, it was pointed out by Zane 
\etal (1995)\markcite{zan95} that most optimistic  models
produce a source density of $\sim 10$ deg$^{-2}$ in ROSAT PSPC 
pointings  at the limiting flux of $10^{-3}$ count
$\rm s^{-1}$, embarrassingly close to the average density of NOIDs,
$\sim 30$ deg$^{-2}$.
A even more dramatic reduction of the number of ONSs could follow from
the recent investigation of ROSAT sources in dark clouds and
high latitude molecular clouds by 
Danner (1996)\markcite{dan96}.  For these fields predictions seem 
to be in excess by a factor $\sim 10-100.$

Although the paucity
of ONSs may be partly related to the heating of the ISM by the source itself
(Blaes, Warren, \&  Madau 1995)\markcite{bwm95},
it could be also closely connected to  the 
properties of NSs at birth and/or  to the long term evolution
of their physical parameters.
Present estimates on the observability of ONSs rely on a number of physical 
hypotheses concerning the velocity distribution of pulsars at birth
and the evolution of the magnetic field (e.g. Treves, \& 
Colpi 1991\markcite{tc91}; Blaes, \& Madau 
1993\markcite{bm93};  Colpi, Campana, \& Treves 1993\markcite{cct93}; 
Zane \etal 1995\markcite{zan95}).
The velocity distribution of ONSs was derived from the evolution, over the 
lifetime of 
the Galaxy, of the pulsar velocity distribution of Narayan, \& Ostriker 
(1990\markcite{no90}, see also Paczy\'nski 1990\markcite{pa90}) 
while  a relic magnetic field $\lesssim 
10^{10}$ G  was assumed a priori. 

Disappointingly, both the velocity distribution of 
NSs and the long--term evolution 
of the magnetic field are still affected by large uncertainties which make
them highly controversial 
issues. The actual number of detectable ONSs depends crucially 
on the number of low velocity neutron stars, which are those accreting at
the highest rate. This, in turn, is directly related to the NS velocity
distribution at birth. Lyne, \& Lorimer 
(1994)\markcite{ll94} and Hansen, \& Phinney (1998)\markcite{hp98} derived 
a pulsar velocity distribution containing fewer slow objects than in
Narayan, \& Ostriker (1990)\markcite{no90}, but this result
is in contrast with other recent investigations 
by Hartman 
(1997)\markcite{ha197} and Hartman \etal (1997)\markcite{ha297}, which indicate
that the number of pulsars with $v\lesssim 40$ kms$^{-1}$ is the same as in 
Narayan \& Ostriker or even higher. 

Observations show that  
neutron stars with magnetic fields $\ll 10^{12}$ G, possibly  $10^9$ G, are 
definitely present in LMXBs and in millisecond pulsars, suggesting  
that the magnetic field have decayed 
in these systems. Instead, no firm conclusions have been
reached yet for isolated objects (see e.g. Srinivasan
1997\markcite{sri97}; Wang 1997\markcite{wa97}). 
Theoretical results are far from being univocal and 
predict either exponential/power--law field decay (Ostriker \& Gunn 
1969\markcite{og69}; Sang \& Chanmugam 1987\markcite{sc87}; 
Goldreich \& Reisenegger 1992\markcite{gr92}; 
Urpin, Chanmugam \& Sang 1994\markcite{urp}; Miri 1996; Urpin, \&
Muslimov 1992) or
little or no decay at all within the age of the Galaxy (Romani 
1990\markcite{ro90}; Srinivasan \etal 1990\markcite{sri90}; 
Goldreich \& Reisenegger 1992\markcite{gr92}; Pethick, \& Sahrling 
1995\markcite{ps95}; see also Lamb 1991\markcite{la91} for a revue). 
Statistical 
analyses based on observations of isolated radio pulsars give equally 
ambiguous results (Narayan \& Ostriker 1990\markcite{no90}; 
Sang \& Chanmugam 1990\markcite{sc90}; 
Bhattacharya \etal 1992\markcite{bha92}), owing in part to the 
difficulty of treating selection effects (Lamb 1992\markcite{la92}). 
At the present, 
there is no clear evidence of field decay during the pulsar phase, but this 
does not preclude the possibility of field decay over longer 
timescales. Different approaches to pulsar
statistics led, independently, to the conclusion that, if the magnetic 
field decays, then it probably does so over a timescale $\sim 100$ Myr, 
well above the characteristic pulsar lifetime (Srinivasan 
1997\markcite{sri97}, Hartman \etal 1997\markcite{ha297}; Lorimer,
Bailes, \& Harrison 1997).

Very recently, however, Wang (1997)\markcite{wa97} 
and Konenkov, \& Popov (1997)\markcite{kp97} noticed that 
a substantial decay of the magnetic field should have occured 
precisely in 
the ONS candidate RX J0720.4-3125.
This object may  have nevertheless a different origin resulting from 
common envelope evolution in a 
binary system. In the last hypothesis
the decay of the $B$--field may have been induced by accretion 
(see e.g. Shibazaki \etal 1989\markcite{shi89}; Romani 1990\markcite{ro90}).

Keeping in mind these uncertainties, the case of RX J0720.4-3125 may provide,
for the first time, evidence for a decay of the 
magnetic field in aging, isolated NSs.
If evolution leaves  a relic field  $\sim 10^8-10^9$ G
a large fraction of the entire NS population is expected
to accrete the ISM, since major spin--down by 
propeller occured over the lifetime of the stars, 
as pointed out by  Blaes, \& Madau (1993)\markcite{bm93}. 
Actually, previous investigations assumed that {\it 
nearly all\/} ONSs accrete from the ISM (see again Treves, \& 
Colpi 1991; Blaes, \& Madau 1993\markcite{bm93}). 
At the low rates typical of ONSs embedded in the ISM, however, 
the conditions under which accretion is possible are rather stringent, since
the inflowing gas may or may not penetrate inside the accretion and the Alfven 
radii according to the star velocity, spin period and 
magnetic field. As a consequence the number of ONSs
accreting today may depend sensibly  on
the distribution of the stellar parameters at birth but,
even more, on the long--term variation  of the spin
and magnetic field.

The aim of this investigation is to explore the role of the magnetic field 
evolution in affecting theoretical predictions for ONSs observability. 
In order to do this, we relax the assumption of a constant, relic 
magnetic field and consider a model in which 
spin evolution causes the core field to migrate to the crust where
dissipation processes (such as electron--phonon scatterings and 
scatterings on impurities)  drive the field  decay 
(see e.g. Srinivasan 
\etal 1990\markcite{sri90}; Urpin, \& Muslimov 1992\markcite{urp92}).
Several evolutionary tracks have been computed and
the properties of the ONS candidate RX J0720.4-3125 are easily
recovered. We show that the paucity of ONSs detections can be accounted for
if the magnetic field decays on a characteristic timescale $\sim 
10^8-10^9$ yr.
The approach presented here suggests that, in close analogy with what is
done for pulsars, the study of ONSs statistics may reveal a promising probe 
for constraining the  spin and 
magnetic field evolution of neutron stars.

\section{Spin and Magnetic Field Evolution}

NSs are widely believed to be borne with high magnetic fields,
$B\approx  10^{12}$ G, and very short rotational  periods $P\approx 10^{-2}$ 
s. During the early phase of the evolution, 
magnetic dipole losses spin down the star at a rate which, in vacuo, is
\begin{equation}
\label{dipoler}
{dP\over dt} \simeq 10^{-8}B_{12}^2\,P^{-1}\, \sy \, , 
\end{equation} 
where $B_{12}$  is the surface dipole field in units of 10$^{12}$ G. 
The NS mass and radius have been assumed to be $M_*=1 M_{\sun}$ and $R_*=10$ km
throughout.
Actually,  isolated NSs are surrounded  by the diffuse interstellar medium, 
so the pulsar relativistic wind continues until the 
ram pressure of the infalling material exceeds the outflowing momentum 
flux at the accretion radius (e.g. Blaes, \& Madau 1993\markcite{bm93}; Treves, 
Colpi, \& Lipunov 1993\markcite{tcl93} and references therein). 
This occurs when the period exceeds a critical value 
\begin{equation}
\label{pcr}
P_{prop} \simeq  10 n^{-1/4} v_{10}^{1/2} B^{1/2}_{12} \, {\rm s} \, , 
\end{equation}
where $v_{10}=(V_{10}^2+c_{s,10}^2)^{1/2}$, $V_{10}$ is the ONS velocity in 
units of 10 km s$^{-1}$, $c_{s}\sim 10 \ {\rm kms}^{-1}$ is the ISM sound 
speed and $n$  the ISM number density in cm$^{-3}$. Here and in the following 
the Bondi formula for the 
accretion rate
\begin{equation}
\label{mdot}
\dot M\simeq 10^{11}n v_{10}^{-3}\, {\rm g\, s^{-1}}
\end{equation} 
is used.
For $P>P_{prop}$ the incoming gas can penetrate  down to the Alfven radius
\begin{equation}
\label{rmag}
R_A\simeq 1.7\times 10^{10} B_{12}^{4/7}\,n^{-2/7}v_{10}^{6/7}\,\,\, \rm 
{cm}\, . 
\end{equation}
The interaction of matter with the  magnetosphere  
prevents, however, accretion to go any further, because, for typical ONSs 
parameters,
the Alfven radius exceeds the corotation radius $R_{co}=1.5\times 10^8
P^{2/3}\,\,\rm {cm}$.
Since  the flow is forced into superkeplerian rotation, angular momentum
will be extracted and 
spin--down proceeds at a rate 
(the propeller phase, Illarionov, \& Sunyaev 1975\markcite{is75})  
 
\begin{equation}
\label{propeller}
{{dP}\over{dt}} \simeq 
10^{-8}n^{9/13}v_{10}^{-27/13}B_{12}^{8/13} P^{21/13}\, \sy.
\end{equation}
Only when the period exceeds 
\begin{equation}
\label{pac}
P_a\simeq 2.5\times 10^{3}\,B_{12}^{6/7} 
n^{-3/4}v_{10}^{1/7}\, {\rm s}.
\end{equation}
steady accretion onto the star surface can set in.  

>From equations (\ref{dipoler}), (\ref{pcr}), (\ref{propeller}) and (\ref{pac}) 
both the critical periods, $P_{prop}$
and $P_a$, as well as the spin--down rates, depend on the magnetic
field. Hence the overall picture of the NS evolution would  
change significantly if the  magnetic field is itself evolving in time. 
In this paper we consider a simple decay model, in which the 
magnetic field is anchored to both the 
superfluid core and the crust, so that its evolution is primarily controlled
by the star rotation (see e.g. Ding, Cheng, \& Chau 1993\markcite{dcc93}; 
Miri, \& Bhattacharya 1994\markcite{mr94}). A different model, in which 
the field is confined in the star crust will be briefly discussed in section
5. In the present picture, the neutron vortices, 
carrying the angular momentum, are pinned to the
proton fluxoids. The radial drift of the vortices, induced by 
spin--down, causes the magnetic 
flux to migrate to the crust where ohmic dissipation occurs (Srinivasan 
1990\markcite{sri90}; Srinivasan 1997\markcite{sri97}).  
The physics of the interaction between vortices and fluxoids 
is still uncertain, but, as shown by 
Ding, Cheng, \& Chau 
(1993)\markcite{dcc93} (see also Miri, \& 
Bhattacharya 1994\markcite{mr94}), the decay rate of the core field 
$B_c$ turns out to be approximately equal to the spin--down rate
\begin{equation}
\label{bcore}
\frac{dB_c}{dt} = - \frac{dP}{dt}\, ,
\end{equation}
yielding $B_c = B_c(0)P(0)/P$. 
Ohmic dissipation takes place in the crust inducing the decay of currents on  
a characteristic timescale $\tau_c$ 
(see again Ding, Cheng, \& Chau 
1993\markcite{dcc93}; Miri, \& Bhattacharya 1994\markcite{mr94}). 
This 
results in the evolution equation for the surface magnetic field $B$
\begin{equation}
\label{bsur}
{{dB}\over{dt}} = -{{B - B_c}\over{\tau_c}}\, .
\end{equation}
The crustal timescale $\tau_c$ depends on the mechanism responsible for
ohmic dissipation and is, in general, a function of time. However, in 
the spirit of the simple model used here and because 
of the still poorly known details of the microphysics, we 
found more convenient to view $\tau_c$ as a constant which we vary 
to bracket uncertainties. From equation 
(\ref{bsur}) one can see that the core and surface fields evolve on the
same timescale if $\tau_c$ is shorter than the spin--down time (controlling
the evolution of $B_c$). In the opposite case, the core field decays 
faster while the surface field relaxes to $B_c$ on the scale $\tau_c$.

\section{Do Aging NSs Accrete Steadily ?}

In this section we follow the evolution of the spin and the surface field 
integrating numerically equations (\ref{dipoler}), (\ref{propeller})
and (\ref{bsur}). At the onset of evolution, at $t=10^3$ yr, the neutron star 
is assumed to spin rapidly with a period  $P(0) = 0.01$ s, while the 
initial value of the surface field is taken to be consistent with the 
narrow gaussian distribution ($\log B_0=12.4$, $\sigma=0.32$) inferred from 
pulsar observations (see e.g. Bhattacharya \etal 1992\markcite{bha92}). 
Accordingly, we restrict our analysis to values of $B(0)$ in the range
$8\times 10^{11}<B(0)<7\times 10^{12}$ G.
The initial core field $B_c(0)$ is taken to be always equal to 
$B(0)/2$ (Ding, Cheng, \& Chau 1993\markcite{dcc93}).

The evolution is divided into three phases (see e.g. Blaes, \& 
Madau 1993\markcite{bm93}; Treves, Colpi, \& Lipunov 1993\markcite{tcl93}; 
Wang 1997\markcite{wa97}). As long as  
$P<P_{prop}$, rotational energy losses are due to
magnetic dipole radiation and the behaviour of $B$ and $P$ is governed by 
equations (\ref{dipoler}) and (\ref{bsur}). For $P_{prop}<P<P_a$, equation 
(\ref{dipoler}) is replaced with (\ref{propeller}) to account for
spin--down by propeller. Since the latter depends on the accretion rate, 
we vary the star velocity $V$ and keep the ISM density constant 
($n=1\, {\rm cm^{-3}}$), in exploring different evolutionary tracks. 
The numerical integration is halted either when  
the period $P$ becomes equal to $P_a$ (calculated with the 
current value of $B$) after a time $t_a$ or when 
$t=t_G=10^{10}$ yr, the age of the Galaxy. As discussed by Wang 
(1997)\markcite{wa97}, during the accretion 
phase the mass loading of the field lines could become important in 
spinning down the star, because it increases the moment of inertia of the 
star/magnetosphere system. However, in the case of RX 
J0720.4-3125 and exponential decay, it only produces a 
small variation (from 7.86 s to 8.39 s) in the value of $P$. Since here we 
focus on the average properties of the 
ONS population, we neglect this further source of spin--down, assuming 
that $P$ (and, accordingly, $B_c$) is a constant for $t>t_a$. The surface 
magnetic field, however, continues to decay and equation (\ref{bsur}) gives
\begin{equation}\label{bacc}
B\left (t \right ) = 
\left (B_c\right )_a + 
\left [B_a - \left (B_c\right )_a \right ]\exp {\left ( - { t - t_a  
\over \tau_c }\right ) } \, ,  
\end{equation}
where $\left (B_c \right )_a= B_c(0)P(0)/P_a$. 
In the following we consider different values of the decay time $\tau_c$
and discuss the ensuing evolutionary scenarios.

\subsection{Very Fast and No Decay}

Although statistical studies of isolated radio pulsars indicate that little 
or no field decay occurs over the pulsar lifetime (Bhattacharya \etal 
1992\markcite{bha92}; Lorimer, Bailes, \& Harrison 1997\markcite{loba97}),
values of $\tau_c\sim 10^7$ yr can not be completely ruled out. 
Anyhow, we find that rapid crustal field decay inhibits angular 
momentum losses to such an extent that virtually no ONS would be accreting 
today, all of them would either be dead pulsars or stay forever in the 
propeller
phase. Quite the opposite happens if $\tau_c=10^{10}$ yr, a value large
enough to mimic the case of no field decay in the population of ONSs. In
this case
most neutron stars enter the accretion stage remaining highly magnetized and
rotating very slowly, with periods up to few days.

\subsection{Decay on Intermediate Timescales}

We start our discussion examining in some detail the evolution of $P$ 
and $B$ for $\tau_c=10^8$ yr. 
Figure 1a, b illustrates the history of a low velocity neutron 
star with $V_{10}=2$ and $\log B(0)=12.4$ G. The endpoint of the
evolution has properties similar to those  of the neutron star in 
RX J0720.4-3125 (Wang 1997\markcite{wa97}). As it can be seen, during 
the lifetime of the Galaxy the star goes through all the three phases, 
pulsar, propeller and 
accretor, and may be visible at present as an accreting source. 
The situation is entirely different for a NS with higher speed, $V_{10}=30$. 
In this case, after the 
initial loss of rotational energy
by magnetic dipole radiation, the star enters the propeller phase, but 
has no time to spin down sufficiently 
to become an accretion--powered X--ray source because $\dot M$ is 
too small to make braking effective. This means that NSs borne with high 
kick velocities may undergo only the pulsar and propeller phases over their
entire lifetime.

Neutron stars which become accretors arrive at the end of the propeller
phase with a period $P_{a}$ and a surface field $B_{a}$ which depend
on the accretion rate, and hence on the star velocity, once $n$, 
$B(0)$ and
$\tau_c$ are fixed. The behaviour of $B_{a}$ and $P_{a}$ as functions of $V$
is shown in figures 2a, b where each curve corresponds 
to a different value of $B(0)$. In figure 2a triangles and squares 
mark the loci where $t_a=10^9$ and $10^{10}$ yr, respectively. 

As it can be seen, for each values of $B(0)$ there exists a limiting
velocity, $V_{cr}$, above which the time required to reach $P_a$ exceeds 
the age of the Galaxy. 
Clearly $V_{cr}$ depends on $B(0)$: NSs with large fields at birth
can reach the accretion phase even if they move with comparatively 
large speed, while for weak initial fields rapidly moving objects remain 
in the propeller stage.
We notice (see figure 2a) that as $V$ increases 
toward $V_{cr}$ substantial field decay has occured before the 
accretion phase begins, and $B_a$ approaches $\sim 10^9$G, irrespective of 
$B(0)$. 
For $V\sim V_{cr}$, $B_a$ represents the relic field, since $\tau_c\ll 
t_a\sim t_G.$
The distribution of the periods is rather wide and ranges from few seconds
to several minutes (see figure 2b). For low values of $B(0)$ 
(bottom curve of figure 2b) $P_a$ monotonically increases with $V$, while 
at larger $B(0),\,$ $P_a(V)$ exhibits a maximum. This is a not in contrast 
with equation (\ref{pac}) because $P_a$ depends also on $B_a$ which, in turn,
is a function of $V$. 

In our model the 
periods at $t=t_a$ coincide with those at $t=t_G$, 
while this is not the case for the surface field because of 
further ohmic dissipation for $t_a<t<t_G$. The 
distribution of the surface field at present, $B_{now}=B(t_G)$, is plotted 
together with the core field in figure 3a. 
As it appears from figure 3a, 
decay is fast enough to freeze the surface fields to the core field at 
$t=t_a$.
ONSs with spin periods $\sim 5-10$ minutes and low 
magnetic field ($B \sim 10^8$ G) can be easily accommodated by the 
model for $\tau_c=10^8$ yr (see figure 3b). 
 
Ohmic dissipation of the crustal field on a scale $\simeq 
10^9$ yr speeds up the evolution to the final state of
accretor. Both losses by magnetic dipole braking and propeller  
slow down the NS effectively.
Consequently, $V_{cr}$ takes higher values and also NSs with large enough 
velocities can enter the accretion stage.
Spin periods vary between few seconds and several  hours and, despite the 
longer crustal decay time, $B_{now}$ may be lower than for $\tau_c=10^8$ 
yr.  This is because 
the larger spin--down rate produces a significant decay of the core field
and there is enough time after the onset of accretion 
for the surface field to approach the core field.
Typical values of $B_{now}$ are between $10^8$ and $10^9$ G (see figure 4a). 
We notice that for $\tau_c=10^9$ yr the period of RX J0720.4-3125 
can not be reproduced along any of the evolutionary tracks (see figure 4b).
 
\section{More On  ONSs Detectability}

Within  our  spin--induced field decay model, 
we now estimate the fraction of NSs (relative to the total number) 
that are presently in the propeller, $N_{prop}$,  and  in the accretor 
phase, $N_{a}$.
To this end, we recall that only those stars having $V>V_{cr}$
($V<V_{cr}$) are in the propeller (accretor) phase.
$N_{prop}$ and $N_{a}$ depend on the 
velocity distribution of ONSs $G(V,t)$, which changes during the evolution of 
the Galaxy. Starting from the pulsar distribution 
of Narayan, \& Ostriker (1990)\markcite{no90},
Blaes, \& Madau (1993)\markcite{bm93} and Zane \etal (1995)\markcite{za95} 
derived  the present distribution 
function following  the evolution of a large number of stars moving 
in the Galactic potential. The current cumulative velocity distribution may
be conveniently approximated as
\begin{equation}
\label{veldis}
G(V,t_G)= \frac{x^m}{1+x^m}
\end{equation}
where $x=V_{10}/6.9$, $m = 3.3$ and $G(\bar V,t_G)$ gives the fraction of 
stars with $V<\bar 
V$. The stars enter the propeller phase at $t\gtrsim 10^8$
yr, so there have been time enough for their velocity distribution to 
evolve. In 
sections 2 and 3, we neglected this effect and assumed that $V$ is
constant in time. We judged this to be a minor flaw
with respect to other sources of uncertainty present in our model, including 
the decay model itself. 

Table 1 contains $N_{a}$ and $N_{prop}$ for different $\tau_c$ and for 
$\log B(0)=12.4$.
Previous models relied on the hypothesis that 
{\it all\/} ONSs are currently accreting, i.e. $N_a = 1$. 
Here we find that
only  for $\tau_c\sim  10^{10}$ yr, most ($\sim 85\%$) of the NSs are now
accreting. This figure  
drops significantly to  $\sim 5\%$  if we allow for crustal field decay
on a timescale $\tau_c\sim10^8$ yr.
However, the fraction of potentially observable NSs is not $N_a$, but 
depends on the characteristics of the detector. Here we refer to ROSAT, 
and the fourth column of table 1  gives  the fraction $N_T$
of accreting NSs  with emission  falling in the T--band (0.1--2.4 keV). The 
energy at 
which a  blackbody spectrum peaks was taken as representative of the 
mean photon energy and  
the effective temperature of radiation was estimated
considering polar cap emission (see figure 5).
It is of interest to notice that, although the fraction of accreting NSs 
increases with $\tau_c$, the relative number of  sources in band is smaller  
for $\tau_c=10^9$ yr. The reason is that the surface field has decayed
more (see figure 4a) than in the case $\tau_c\sim 10^8$ yr, so the emitted 
spectra are softer. We find that,  
for $\tau_c\sim 10^{8-9}$ yr, $N_T$ clusters around $\sim 4\%$.
Yet, this number can not be compared with previous figures
on the number of detectable ONSs (see Treves, \& Colpi 1991\markcite{tc91};
Zane \etal 1995\markcite{za95}) 
since they evaluated the number of sources with a flux above the sensitivity 
threshold of ROSAT. In order to obtain a meaningful comparison, we have 
computed 
the  fraction $N_T^{old}$ of ONSs in the T--band, assuming a magnetic
field of $10^9$ G,  $n=1$ cm$^{-3}$ and blackbody polar cap emission.
$N_T^{old} = 0.21$ gives precisely the number of sources in the T--band
in the spirit of previous models and  $R = N_T/N_T^{old}$
is the ratio between the present and past predicted numbers.
The values of $R$ are given in Table 1 and  indicate that the expected 
number is  reduced  by a factor $\sim 4$ for $\tau_c=10^8$ yr.
$R$ is even smaller for $\tau_c=10^9$ yr owing to the more severe
decay of the core field. 
For $\tau_c\sim 10^7$ yr, $R$ is as low as $0.03$; 
by contrast, $R$ exceeds unity for $\tau_c=10^{10}$ 
yr. We note that $N_{prop}$ and $N_a$ are only very weakly 
dependent on the initial value of the period: models with larger
values of $P(0)$ give essentially the same figures.

\section{Discussion and Conclusions}

In this paper we have shown that the expected number of ONSs  
observable
with ROSAT can be sensibly reduced if the star magnetic field decays
on timescales $\sim 10^8-10^9$ yr. \footnote{Soon after this paper was
submitted, we became aware that a similar conclusion was reached by Livio,
Xu, \& Frank (1998)\markcite{lxf98}.} This
is primarily due to the
lower
fraction of NSs which are in the accretion phase now, but also
to a softening of the spectra which may bring the
peak of the emission outside ROSAT T--band if $\tau_c=10^9$ yr. 
It follows that, if
the paucity of detected ONSs is to be explained in terms of a global
reduction in the number of accretion--powered sources, 
decay times $\tau_c\lesssim 10^8$ yr, should be invoked.
For intermediate values of $\tau_c$, $\sim 10^9$ yr,  the 
fraction of accreting ONSs is $\sim 50-60\%$, too large to match
observational limits, but only $\sim 15\%$ of the sources falls in
the ROSAT T--band, yielding again an acceptable number of 
{\it detectable\/} objects. This last figure should be regarded as indicative,
being based on the assumption that all sources emit a blackbody spectrum
from the polar caps.
If the decay time is
$\sim 10^{10}$ yr or larger, spin--down by propeller is so effective
that nearly all ONSs are accreting today. In this case the surface 
magnetic field is close to the initial value, $B\sim 10^{12}$ G,
and deviations from a pure blackbody spectrum are to be expected.
Unfortunately, no detailed radiative transfer calculations are
available for accretion atmospheres onto magnetized NSs, at least
to our present knowledge. Some hints may be, nevertheless, derived
from the analysis of radiative transfer in cooling neutron stars
atmospheres (Shibanov \etal 1992\markcite{shi92}), 
since the input physics of the two model is quite 
similar. The emerging picture indicates that 
magnetic effects tend, somehow, to balance the hardening found by
Zampieri \etal (1995)\markcite{ztzt95} in the case of unmagnetized accretion 
atmospheres, so 
that, for $B\sim 10^{12}$ G, spectra are, globally, not too different from a 
blackbody emitted by the polar caps. With this proviso, figures
in table 1 tell us that all accreting sources should be indeed within
ROSAT T--bandpass and show that decay times 
$\tau_c\gtrsim 10^{10}$ yr provide a fraction of potentially detectable ONSs
equal to that derived in previous investigations. 

Our main conclusion is that models of spin--induced magnetic field decay with a 
crustal dissipation time shorter than the age of the Galaxy are able to bring 
theoretical predictions of ONSs detectability closer to present
observational limits. It is interesting to note that ohmic 
diffusion times between $10^{8.5}$ and $10^9$ yr
can also explain the low values of $B$ observed in millisecond pulsars, 
which experienced spin induced field decay during  binary evolution
in wide low--mass systems (Miri, \& Bhattacharya 1994\markcite{mb94}).

Although our present results seem indeed
to point to a decay of magnetic field in ONSs, a firmer conclusion 
may be reached once the number of sources actually present in ROSAT fields
is computed evolving the ONSs distribution function together with the
spin and the magnetic field. Two other issues deserve further discussion.
In computing the spin--down rate in the propeller phase the original 
expression of Illarionov, \& Sunyaev (1975)\markcite{is75}
was used. It should be noticed that the physics governing the propeller
is rather complex and still far from a complete understanding. It has
been shown (see e.g. Ghosh 1995\markcite{go95}) that the characteristic
time for angular momentum losses strongly depends on the details of the
accretion process and can be longer than what is assumed here. If the 
propeller
is less efficient in braking the star, the number of accreting ONSs 
reduces further. On the other hand numerical simulations (Wang, \& Robertson 
1985\markcite{wr85}) seem to indicate that propeller spin--down can go even 
faster that in equation (\ref{propeller}). 

The second point concerns
the field decay model. If the field is confined only in the stellar
crust (see e.g. Urpin, \& Muslimov 1992\markcite{urp92}) 
the emerging picture may change 
since the magnetic field is not coupled to the period.
In the model of Urpin, \& Muslimov $\tau_c$  is computed from the 
microphysics and varies in time, depending on the thermal evolution 
of the stellar crust.
We have repeated the calculation using 
equation (\ref{bsur}) with  $B_c=0$ and $\tau_c$ estimated from 
figure 5 of Urpin, \& 
Muslimov,  
and found that the number of accreting ONSs again reduces. Even
if the initial surface field is as large as $10^{13}$ G only $\lesssim 1\%$
of the NS population is the accretion phase now. 
On the opposite, an exponential decay of the surface field with 
$\tau_c\gtrsim 10^9$ yr, as  inferred  by 
Pethick, \& Sahrling (1995)  in the case of a non--vanishing core field, 
do not seem to solve the problem of the excess of potentially visible ONSs.

The conclusion that many ONSs should be in the propeller stage 
seem rather robust if the magnetic field decays over
a characteristic $\tau_c\sim 10^{9}$ yr, irrespective of the details
of the model.
Is there a way to observe directly these objects ?
The power associated to the propeller derives ultimately from the 
rotational energy, and therefore one can deduce that it should be smaller
than that released  by accretion of matter onto the star surface
typically by a factor $\sim 100$ 
(see e.g. Treves, Colpi, \& Lipunov 1993\markcite{tcl93}). The accreted mass,
however, could accumulate at the Alfven radius until the pressure 
required to overcome the centrifugal barrier is built up, giving raise to 
recurrent episodes of accretion. If this is the case, ONSs in the 
propeller phase could be observed as transient sources (Treves, Colpi, \& 
Lipunov 1993\markcite{tcl93}). 
More definite 
predictions about the observational appearance of these sources, including
spectral properties, luminosity, recurrence times, demand, however, for
a more realistic modelling of the physics governing the propeller.

\acknowledgments
We are grateful  to  Sergio Campana for constructive comments during the 
early stages of this work and to Piero Madau and Andrea Possenti 
for valuable discussions.

\clearpage

\figcaption[fig1.ps]{a) The evolution of the surface field for
$V_{10}=2$, $\log B(0)=12.4$ G and $\tau_c=10^8$ yr. b) The evolution 
of the period 
for the same choice of the parameters; the periods for the onset of the 
propeller (eq. [\ref{pcr}], dashed line)
and of accretion (eq. [\ref{pac}], dash--dotted line) are also shown.
\label{fig1a}}

\figcaption[fig2.ps]{a) The surface field at the onset of accretion as a 
function of the star velocity for
$\tau_c=10^8$ yr, from bottom to top: 
$B(0)=0.87,1.37,2.06,3.06,4.60,7.28\times 10^{12}$ G; triangles and squares
mark the loci where $t_a=10^9$  and $t_a=10^{10}$ yr respectively.
b) Same for the spin period.
\label{fig2}}

\figcaption[fig3.ps]{a) The the surface (full lines) 
and the core field (dashed lines) at present vs. $V$ ($B_c$ and $B$
actually coincide so the two curves are superimposed); details as in
figure 2.
b) The value of the surface field at present vs. the period; the vertical 
line marks the period of RX J0720.4-3125.
\label{fig3}}

\figcaption[fig4.ps]{Same as in figure 3 for $\tau_c=10^9$ yr.\label{fig4}}

\figcaption[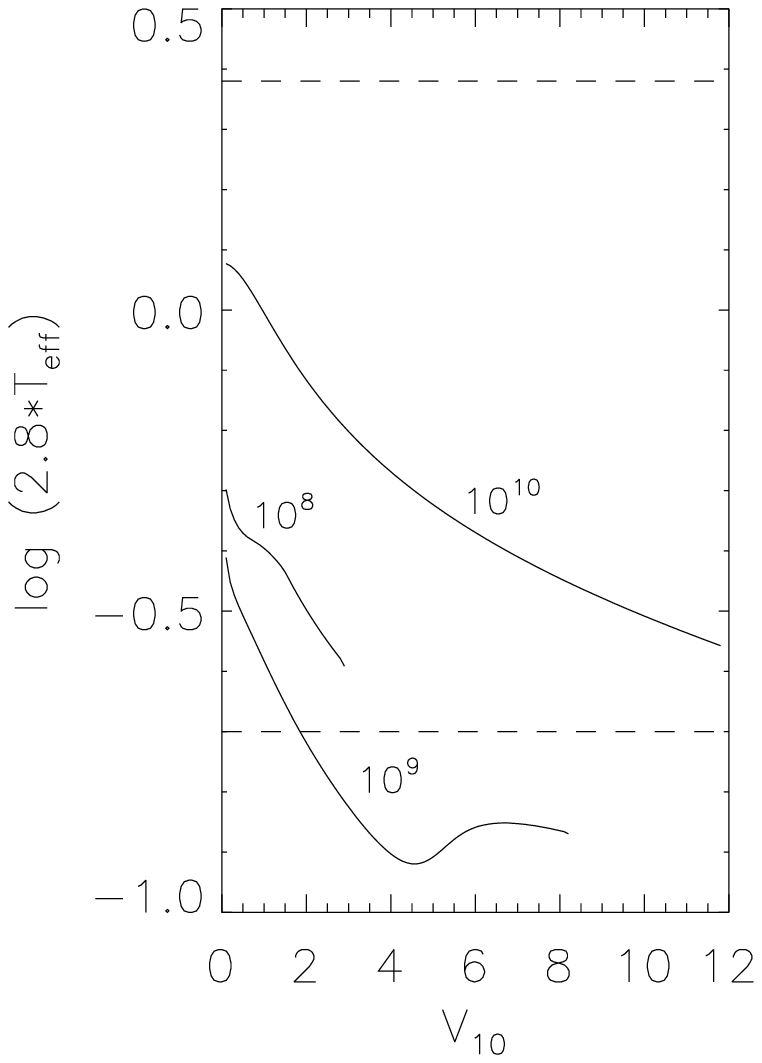]{The effective temperature against the star velocity
for $\log B(0) = 12.4$ and different values of $\tau_c$; the two
horizontal lines mark the ROSAT T--bandpass. \label{fig5}}

\clearpage


\begin{deluxetable}{ccccc}
\tablecolumns{5}
\tablewidth{0pt}
\tablecaption{Fraction of objects in the different 
phases\tablenotemark{\ast}\label{tab1}} 
\tablenum{1}
\tablehead{
\colhead{ $\tau_c$ (yr)} &
\colhead{$N_{prop}$\tablenotemark{a}} &
\colhead{$N_{acc}$\tablenotemark{b}} &
\colhead{$N_T$\tablenotemark{c}} & 
\colhead{$R$}\tablenotemark{d} \nl}
\startdata 
  $10^8$ & $0.95$ & $0.05$ 
  & $0.05$ & 0.23 \\ 
  $10^9$ & $0.37$ & $0.63$ 
  & $0.03$ & 0.16 \\ 
  $10^{10}$ & $0.15$ & $0.85$ 
  & $0.85$ & 4.10 \\ 
\enddata
\tablenotetext{\ast}{Here $\log B(0)=12.4$}
\tablenotetext{a}{Fraction of NSs in the propeller phase}
\tablenotetext{b}{Fraction of NSs in the accretion phase}
\tablenotetext{c}{Sources in the ROSAT 
T--band} 
\tablenotetext{d}{Ratio of sources in the T--band according to present
and past models}
\end{deluxetable}

\end{document}